\documentclass[lineno]{jfm}

\usepackage{graphicx}
\usepackage{newtxtext}
\usepackage{newtxmath}
\usepackage{natbib}
\usepackage{hyperref}
\hypersetup{
    colorlinks = true,
    urlcolor   = blue,
    citecolor  = black,
}

\newcommand{\RomanNumeralCaps}[1]

\title{Passive cell body plays active roles in microalgal swimming via nonreciprocal interactions}

\author{Xiaoping Hu\aff{1},
  Zhaorong Liu\aff{2},
  Da Wei\aff{3}
 \and Shiyuan Hu\aff{1}
 \corresp{\email{shiyuanhu@buaa.edu.cn}}}

\affiliation{\aff{1}School of Physics, Beihang University, Beijing 100191, China
\aff{2}Department of Physics, University of Science and Technology of China, Hefei, Anhui 230026, China
\aff{3}Beijing National Laboratory for Condensed Matter Physics, Institute of Physics, Chinese Academy of Sciences, Beijing 100190, China}

\begin{document}
\maketitle

\begin{abstract}
The cell body of flagellated microalgae is commonly considered to act merely as a passive load during swimming, and a larger body size would simply reduce the speed. In this work, we use numerical simulations based on a boundary element method to investigate the effect of body-flagella hydrodynamic interactions (HIs) on the swimming performance of the biflagellate, \textit{Chlamydomonas reinhardtii}. We find that body-flagella HIs significantly enhance the swimming speed and efficiency. As the body size increases, the competition between the enhanced HIs and the increased viscous drag leads to an optimal body size for swimming. Based on the simplified three-sphere model, we further demonstrate that the enhancement by body-flagella HIs arises from an effective nonreciprocity: the body affects the flagella more strongly during the power stroke, while the flagella affect the body more strongly during the recovery stroke. Our results have implications for both microalgal swimming and laboratory designs of biohybrid microrobots.
\end{abstract}



\section{Introduction}
The unicellular green alga \textit{Chlamydomonas reinhardtii} has been widely used as a model organism for studying various problems in fluid dynamics and cell biology~\citep{Goldstein2015}, such as
tactic behaviors~\citep{Leptos2023}, flagellar dynamics~\citep{Quaranta2015,Sartori2016}, and single cell flow fields~\citep{Drescher2010,Wei2019,Wei2021}. The wild-type cell of \textit{C. reinhardtii} has two flagella anchored at the anterior side of the spheroidal cell body. During swimming, the two flagella beat approximately in a two-dimensional plane, with a breaststroke-like gait consisting of distinct power and recovery strokes. It is now well established that flagellar beating is driven by distributed molecular motors, a characteristic conserved across eukaryotic cells. Studies of the swimming mechanisms of green algae have inspired numerous designs of artificial flagella and microswimmers~\citep[see e.g.,][]{Diaz2021,Moreau2024}.

The analysis of eukaryotic flagellar dynamics often assumes a fixed flagellar base in space. \cite{Machin1958} solved the bending waves of an elastic beam actuated at the boundary. Although this boundary-driven assumption is a significant simplification of realistic flagellar dynamics, later developments reveal the effects of various physical elements on flagellar propulsion, such as effective flexibility~\citep{Wiggins98,Yu2006,Peng17}, swimmer geometry~\citep{Lauga07}, flagellar curvature~\citep{Liu20,Hu22}, and hydrodynamic interactions (HIs)~\citep{Elfasi18,Hu24}. On the other hand, studies that combine mechanical models of flagellar deformation with molecular motor dynamics focus on how beating waveforms are regulated~\citep[see the review by][]{Gilpin20}.

Under dynamic swimming conditions, the modeling of microorganisms involves different complexities, especially in the coupling between the cell body and flagella. In addition to mechanical coupling at the flagellar base, flagella are also coupled to the cell body through long-range HIs. Therefore, combining flagellar beating with a moving cell body is crucial for understanding the swimming mechanisms. Simulations and experimental measurements on various microorganisms have reported a decrease in swimming speed with increasing body size, including spermatozoa~\citep{Higdon79}, monotrichous bacteria~\citep{Kamdar23,Liu2025}, and the ciliate \textit{Volvox}~\citep{Omori2020}. In particular, body-flagella HIs have been shown to influence both the swimming speed and efficiency of bacteria~\citep{Liu2025}. For flagellated microalgae, we expect more complex and stronger body-flagella HIs due to significantly time-varying and closer spacing between them. Over a beating period, the cell body moves back and forth, entraining a considerable volume of fluid. At low Reynolds number (\Rey), this body-induced flow decays slowly in space and can significantly influence flagellar propulsion. Meanwhile, the flow induced by the flagella also modifies the drag force acting on the cell body. As a minimal swimming model, the three-sphere model, in which the flagellar beating is represented by spheres moving along closed orbits, provides insights into flagellar synchronization~\citep{Friedrich2012}, run-and-tumble behavior~\citep{Bennett2013}, and three-dimensional helical trajectories~\citep{Cortese2021}. Optimal gaits for swimming and feeding have been identified using numerical optimization~\citep{Tam2011}, where the swimming dynamics is modeled via flow singularities placed along the flagella and at the center of the cell body. Although these studies reveal different dynamic behaviors of microalgal swimming, the effect of body-flagella HIs on swimming performance remains elusive. In recent work~\citep{Hu24}, we modeled the swimming dynamics of multiflagellated microalgae and demonstrated that body-flagella HIs determine the hydrodynamic advantage of multiflagellarity. In particular, swimming efficiency increases monotonically with the number of flagella only when they are attached to the anterior side of the cell body. However, whether an optimal body size exists for microalgal swimming is still an open question. 

In this work, we use numerical simulations to investigate the effect of body-flagella HIs on the swimming performance of \textit{C. reinhardtii}. We prescribe the flagellar waveform based on experimental measurements and incorporate both body-flagella and interflagellar HIs using a boundary element method. We find that body-flagella HIs can significantly enhance the swimming performance and lead to an optimal body size for both speed and efficiency. Further analysis based on the three-sphere model reveals that body-flagella HIs are effectively nonreciprocal. 

\section{Simulation model}\label{model}
\begin{figure}
\centerline{\includegraphics{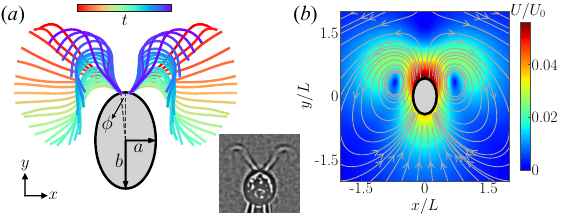}}
\caption{Model of swimming \textit{C. reinhardtii}. (\textit{a}) Flagellar waveform measured from experiments. Blue indicates recovery stroke, and red indicates power stroke. (\textit{b}) Time-averaged disturbance flow field generated by the swimming model.}
\label{fig1}
\end{figure}

Our swimming model of \textit{C. reinhardtii} consists of two flagella of length $L$ and a spheroidal cell body with semi-minor axis $a$ and semi-major axis $b$. Planar waveforms of cells held stationary are measured experimentally via high-speed videography (figure~\ref{fig1}\textit{a}) and serves as input for numerical simulations. Physical parameters of a sample cell are given in table~\ref{table1}. We parameterize each flagellum by the arclength $s\in [0,L]$ along its centerline. From the measured waveform, we evaluate the velocities of the two flagella relative to the cell body, $\boldsymbol{U}_i(s,t)$ for $i = 1,2$, at each recorded time instant. In this work, we focus on the normal breaststroke gait. Although a small phase lag (approximately $0.1\pi$) exists between the \textit{cis} and \textit{trans} flagella~\citep{Wan2014}, only a sufficiently large phase difference ($\gtrsim \pi/2$) can lead to a non-negligible rotation of the cell body~\citep{Geyer2013}. Therefore, the two flagella are assumed to beat mirror-symmetrically, and the swimming model undergoes only translational motion. We further neglect the non-planar components of flagellar beating, as the resulting rotation frequency of the cell is much smaller than the flagellar beating frequency~\citep{Cortese2021}.
\begin{table}
\begin{center}
\begin{tabular}{ccc}
   Parameter & Symbol  & Value \\
   \hline
   beating frequency & $f_0$ & 49 Hz \\ 
   flagellum radius, length  & $r_0$, $L$ & 0.125, 14 $\mu$m \\
   anchored angle & $\phi$ & 0.05$\pi$ \\
   semi-minor, semi-major axes & $a$, $b$ & 3.89, 5.95 $\mu$m \\
\end{tabular}
\caption{Parameters of a representative \textit{C. reinhardtii} cell measured in experiments. The value of flagellum radius is taken from~\cite{Sager1957}.}
\label{table1}
\end{center}
\end{table}

We simulate the dynamics of the model swimmer using a hybrid boundary element and regularized Stokeslet method~\citep{Smith2009}. The regularized Stokeslet is the exact solution to the Stokes equation subjected to a smoothed point force. With a specific smooth function, it can be expressed as~\citep{Cortez2005}
\begin{equation}\label{reg_stokeselt}
\mathsfbi{G}_{\epsilon} (\boldsymbol{x}_0, \boldsymbol{x}) = \frac{\mathsfbi{I}(R^2 + 2 \epsilon^2) + \boldsymbol{R}\boldsymbol{R}}{R_{\epsilon}^3},
\end{equation}
where $\epsilon$ is the regularization parameter, $\boldsymbol{R} = \boldsymbol{x}_0-\boldsymbol{x}$, and $R_{\epsilon} = \sqrt{R^2 + \epsilon^2}$. 

We represent the motions of the cell body and the flagella by distributions of regularized Stokeslets on the body surface and along the flagellar centerlines, respectively. As verified by a test problem on a rigid scallop in Appendix~\ref{appA}, this flagellar representation yields swimming dynamics that closely matches that obtained from a non-local slender body theory~\citep{Johnson1980}. Denote the body surface by $S$ and the flagella centerlines by $C_i$ with $i=1,2$. The disturbance velocity in the fluid domain generated by the motion of the cell body is given by
\begin{equation}\label{body_velo}
\boldsymbol{u}_{\mathrm{b}}(\boldsymbol{x}_0) = \frac{1}{8\pi\mu} \int_{S} \mathsfbi{G}_{\epsilon}(\boldsymbol{x}_0, \boldsymbol{x}) \bcdot \boldsymbol{f}_{\mathrm{b}}(\boldsymbol{x})\, \mathrm{d}\boldsymbol{x}, 
\end{equation}
and that by the flagella is given by
\begin{equation}\label{flagella_velo}
\boldsymbol{u}_{i}(\boldsymbol{x}_0) = \frac{1}{8\pi\mu} \int_{C_i} \mathsfbi{G}_{\epsilon}[\boldsymbol{x}_0, \boldsymbol{x}_i(s)] \bcdot \boldsymbol{f}_{i}(s)\, \mathrm{d}s, \quad \text{for }i = 1,2, 
\end{equation}
where $\mu$ is the fluid viscosity, $\boldsymbol{x}_i(s)$ is the centerline position, and $\boldsymbol{f}_{\mathrm{b}}$ and $\boldsymbol{f}_{i}$ are the hydrodynamic force densities exerted by the body and flagella on the fluid, respectively. When the evaluation points $\boldsymbol{x}_0$ lie on the boundaries, the total disturbance velocity due to the flagella and cell body satisfies the following constraints imposed by the no-slip condition:
\begin{gather}
\boldsymbol{u}_\mathrm{b}(\boldsymbol{x}_0) + \sum_{i} \boldsymbol{u}_{i}(\boldsymbol{x}_0) = \boldsymbol{U}_{\mathrm{b}}(t), \quad \text{for } \boldsymbol{x}_0 \in {S},\label{noslip1} \\
\boldsymbol{u}_\mathrm{b}(\boldsymbol{x}_0) + \sum_{j} \boldsymbol{u}_{j}(\boldsymbol{x}_0) = \boldsymbol{U}_{\mathrm{b}}(t) + \boldsymbol{U}_i(s,t), \quad \text{for }\boldsymbol{x}_0 \in {C_i} \text{ and }i,j = 1,2, \label{noslip2}
\end{gather}
where $\boldsymbol{U}_{\mathrm{b}}(t)$ is the instantaneous velocity of the cell body. Equations~(\ref{noslip1}) and (\ref{noslip2}) incorporate both the interflagellar and the body-flagella HIs. Finally, the total hydrodynamic force and torque are zero:
\begin{gather}
\int_{S}\boldsymbol{f}_{\mathrm{b}}(\boldsymbol{x})\, \mathrm{d}\boldsymbol{x} + \sum_i \int_{C_i} \boldsymbol{f}_i(s) \, \mathrm{d}s = 0, \label{force_free}\\
\int_{S} (\boldsymbol{x}-\boldsymbol{x}_{\mathrm{b}}) \times \boldsymbol{f}_{\mathrm{b}}(\boldsymbol{x})\, \mathrm{d}\boldsymbol{x} + \sum_i \int_{C_i} [\boldsymbol{x}_i(s)-\boldsymbol{x}_{\mathrm{b}}] \times \boldsymbol{f}_i(s) \, \mathrm{d}s = 0, \label{torque_free}
\end{gather}
where the torques are evaluated relative to the body center $\boldsymbol{x}_\mathrm{b}$.

We scale lengths by $L$, time by the beating period $T$, velocities by $U_0 = L/T$, and forces by $F_0 = \mu L^2/T$. The body aspect ratio $b/a$ is kept constant when varying the semi-major axis $b$. The integral formulations (\ref{body_velo})--(\ref{torque_free}) are discretized into a system of linear equations for the unknowns $\boldsymbol{U}_\mathrm{b}$, $\boldsymbol{f}_\mathrm{b}$, and $\boldsymbol{f}_i$. Since the Stokeslet $\mathsfbi{G}_{\epsilon}(\boldsymbol{x}_0, \boldsymbol{x})$ often varies more rapidly near the source position $\boldsymbol{x}$ than the hydrodynamic forces, the evaluation of the Stokeslet is performed on a finer grid than that of the forces~\citep{Smith2009}. Equations (\ref{body_velo}) and (\ref{flagella_velo}) are approximated as
\begin{equation}
\boldsymbol{u} (\boldsymbol{x}_0) = \frac{1}{8\pi\mu} \sum_{n=1}^N \boldsymbol{f}^n\bcdot \int_{\delta S^n (\delta C^n)}  \mathsfbi{G}_{\epsilon}(\boldsymbol{x}_0, \boldsymbol{x}) \,\mathrm{d}\boldsymbol{x},
\end{equation}
where $N$ is the number of discrete surface elements $\delta S^n$ (line elements $\delta C^n$) on the cell body (on each flagellum) with force density $\boldsymbol{f}^n$. The integral of $\mathsfbi{G}_{\epsilon}$ over each element is evaluated using Gauss–Legendre quadrature with $m\times m$ points on $\delta S^n$ ($m$ points on $\delta C^n$). For most simulations, we use $N \approx 200$ on the sphere, $N = 25$ on each flagellum, and $m=6$. The regularization parameter $\epsilon$ is taken to be the same as the flagellum radius $r_0$ (table~\ref{table1}). 

\section{Results and discussion}
\subsection{Optimal cell body size for swimming}
Figure~\ref{fig1}(\textit{b}) shows the disturbance velocity field of the model swimmer averaged over a beating period. There is one vortex on each side of the cell body, and a flow stagnation point exists in front of the swimmer along its swimming direction. These features closely resemble the measured flow field around free-swimming cells~\citep{Drescher2010}. Using the parameters of the sample cell (table~\ref{table1}), we compute the time-averaged swimming speed $\langle U_\mathrm{b} \rangle \approx 70.4$ $\mu$m/s, in agreement with the experimentally measured speed~\citep{Buchner2021}. Slight variations in the anchored angle $\phi$ of the flagella (figure~\ref{fig1}\textit{a}) have little effect on the swimming performance. 
\begin{figure}
\centerline{\includegraphics{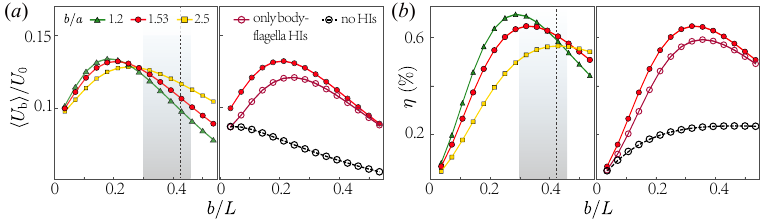}}
\caption{Effect of cell body size on swimming performance: (\textit{a}) $\langle U_\mathrm{b}\rangle/U_{\mathrm{0}}$ as a function of $b/L$; (\textit{b}) swimming efficiency $\eta$ as a function of $b/L$. The shaded areas indicate the distribution of $b/L$ from experimental measurements. The vertical dashed lines mark the value of $b/L$ of the sample cell. Other parameters are given in table~\ref{table1}.}
\label{fig2}
\end{figure}

We first evaluate the dependence of $\langle U_\mathrm{b} \rangle$ on cell body size by varying the semi-major axis length $b$ while using the same flagellar waveform data. As shown in figure~\ref{fig2}(\textit{a}), $\langle U_\mathrm{b} \rangle$ varies non-monotonically with $b/L$, the ratio of body size to flagellar length. As the cell body becomes more slender (larger values of $b/a$), the maximum of $\langle U_\mathrm{b} \rangle$ shifts to larger values of $b/L$. From a collection of approximately 40 cells, we measure an average $b/L = 0.38 \pm 0.08$, as indicated by the shaded regions in figure~\ref{fig2}. For the sample cell ($b/a = 1.53$), the maximum occurs at $b/L \approx 0.21$, which is lower than its actual geometric ratio (0.425; see table~\ref{table1}) and lies more than one standard deviation below the measured average.

We further compute the swimming efficiency $\eta$ using a definition similar to that in~\cite{Lighthill52}:
\begin{equation}\label{efficiency}
\eta = \frac{\zeta_\mathrm{b} \langle U_\mathrm{b} \rangle^2}{\left\langle \int_{S} \boldsymbol{f}_{\mathrm{b}} \bcdot \boldsymbol{U}_{\mathrm{b}} \, \mathrm{d}\boldsymbol{x} \right\rangle + \sum_i \left\langle \int_{C_i} \boldsymbol{f}_i \bcdot (\boldsymbol{U}_{\mathrm{b}} + \boldsymbol{U}_i)\,\mathrm{d}s \right\rangle},
\end{equation}
where $\zeta_\mathrm{b}$ is the friction coefficient of a spheroid translating parallel to its long axis~\citep{Kim13}. Equation~(\ref{efficiency}) expresses the ratio of the work required to drag a spheroid moving steadily at speed $\langle U_\mathrm{b} \rangle$ to the work dissipated by the motions of the flagella and cell body. Figure~\ref{fig2}(\textit{b}) shows that $\eta$ also exhibits a pronounced optimum with respect to $b/L$. Unlike $\langle U_\mathrm{b} \rangle$, for the sample cell, the optimal value of $b/L$ for $\eta$ lies within one standard deviation of the measured average. 

Removing the interflagellar HIs, which amounts to neglecting the disturbance velocity from the other flagellum in (\ref{noslip2}), has only a minor effect on the swimming performance (see the right-hand images in figure~\ref{fig2}\textit{a},\textit{b}). This is due to hydrodynamic screening by the cell body. Further removing the body-flagella HIs, i.e., neglecting $\boldsymbol{u}_i$ in (\ref{noslip1}) and $\boldsymbol{u}_\mathrm{b}$ in (\ref{noslip2}), results in a much lower $\langle U_\mathrm{b} \rangle$ and a monotonically decreasing trend with $b$. Therefore, the surprising increase of $\langle U_\mathrm{b} \rangle$ at small values of $b$ is attributed to increasingly strong body-flagella HIs as $b$ increases. Meanwhile, the intrinsic viscous drag of the cell body $\zeta_\mathrm{b} U_\mathrm{b}$, also increases with $b$ and eventually overwhelms the flagellar propulsion, leading to a decrease in $\langle U_\mathrm{b} \rangle$ at sufficiently large $b$.
\begin{figure}
\centerline{\includegraphics{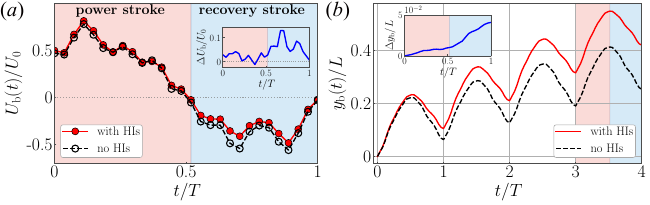}}
\caption{Effect of body-flagella HIs on (\textit{a}) instantaneous swimming speed $U_\mathrm{b}(t)$ and (\textit{b}) instantaneous displacement $y_\mathrm{b}(t)$. Insets show the differences between the results obtained with and without body–flagella HIs for (\textit{a}) $U_\mathrm{b}(t)$ and (\textit{b}) $y_\mathrm{b}(t)$ over a beating period. Red and blue regions indicate power and recovery strokes, respectively.}
\label{fig3}
\end{figure}

In figure~\ref{fig3}, we compare the swimming speed $U_\mathrm{b}(t)$ and the instantaneous displacement $y_\mathrm{b}(t)$ of the cell body, obtained with and without body-flagella HIs. Our key observation is that body-flagella HIs not only increase the forward speed during the power stroke, but also reduce the backward speed during the recovery stroke (figure~\ref{fig3}\textit{a}); as a result, the swimmer achieves a larger forward displacement during the power stroke, and a smaller backward displacement during the recovery stroke (figure~\ref{fig3}\textit{b}). Over one period, the swimmer accumulates a net displacement of approximately 0.1 $L$, which is approximately twice that achieved without body-flagella HIs. 

The appearance of an optimal body size for swimming is robust and not sensitive to variations in flagellar waveforms. In Appendix~\ref{appB}, we present additional simulation results using waveforms measured from different sample cells. Despite large variations in the static curvature, wavelength, and amplitude, the swimming performance consistently varies non-monotonically with $b/L$. This suggests that body-flagella HIs---reflecting how they influence each other’s drag forces---are predominantly determined by their relative, large-scale approaching and receding motions on the order of $L$ during the power and recovery strokes. The robustness also implies that our results are not likely to be affected by slight flagellar adaptations \textit{in vivo} arising from flexibility and active forcing.

\subsection{Three-sphere model of Chlamydomonas swimming}\label{three_sphere}
To reveal the mechanisms underlying the enhancement of swimming performance by body-flagella HIs, we consider a three-sphere model consisting of a central sphere of radius $r_\mathrm{b}$, representing the cell body, and two side spheres of radius $r_{\mathrm{f}}$, representing the flagella (figure~\ref{fig4}\textit{a}). The flagellar spheres are assumed to be much smaller than the cell body, $r_{\mathrm{f}} \ll r_\mathrm{b}$. Unlike previous works~\citep{Friedrich2012, Bennett2013}, we explicitly account for the no-slip boundary condition on the body surface by using the spherical image system $\mathsfbi{G}_{\mathrm{im}}$~\citep{Kim13}. The two flagellar spheres, located at $\boldsymbol{x}_i$ ($i = 1,2$), are prescribed to move with velocity $v_0 \hat{\boldsymbol{t}}_i$ relative to the body and are mirror-symmetric with respect to the model centerline. Due to the force-free condition, the cell body, located at $\boldsymbol{x}_\mathrm{b}$, develops an instantaneous translational velocity $\boldsymbol{U}_{\mathrm{b}}$. The motion of the cell body induces a disturbance flow $\boldsymbol{v}_\mathrm{b}$ in the surrounding fluid,
\begin{equation}
\boldsymbol{v}_\mathrm{b}(\boldsymbol{x}_0, \boldsymbol{x}_{\mathrm{b}}; \boldsymbol{U}_{\mathrm{b}}, r_\mathrm{b}) = \frac{3}{4} r_\mathrm{b} \boldsymbol{U}_{\mathrm{b}} \bcdot \left(1 + \frac{r_\mathrm{b}^2}{6}\nabla^2 \right) \mathsfbi{G}(\boldsymbol{x}_0, \boldsymbol{x}_{\mathrm{b}}),
\end{equation}
where $\mathsfbi{G}$ denotes the free-space Stokeslet. 

The total drag force on the flagellar spheres is $\boldsymbol{F}_i = -6\pi\mu r_\mathrm{f} (v_0 \hat{\boldsymbol{t}}_i + \boldsymbol{U}_\mathrm{b}) + \boldsymbol{F}_{\mathrm{b} \to i}$, where $\boldsymbol{F}_{\mathrm{b}\to i}$ is the contribution to $\boldsymbol{F}_i$ by the cell body, i.e., the body-to-flagella (B-to-F) force, given by
\begin{equation}\label{b_to_f}
\boldsymbol{F}_{\mathrm{b}\to i} = 6\pi\mu r_\mathrm{f} \left[\boldsymbol{v}_\mathrm{b}(\boldsymbol{x}_i, \boldsymbol{x}_{\mathrm{b}}; \boldsymbol{U}_{\mathrm{b}}, r_\mathrm{b}) + \frac{1}{8\pi\mu} \mathsfbi{G}_{\mathrm{im}}(\boldsymbol{x}_i,\boldsymbol{x}_{\mathrm{b}}; r_\mathrm{b})\bcdot \boldsymbol{F}_i \right].
\end{equation}
In (\ref{b_to_f}), $\boldsymbol{v}_\mathrm{b}$ represents the dynamic component of the B-to-F force, and exists only when the cell body is in motion. The term involving $\mathsfbi{G}_{\mathrm{im}}$ represents the static component of the B-to-F force, and captures the effect of a nearby no-slip boundary on the flagellar force. The total drag force on the cell body is $\boldsymbol{F}_\mathrm{b} = -6\pi\mu r_\mathrm{b} \boldsymbol{U}_\mathrm{b} + \sum_i \boldsymbol{F}_{i \to \mathrm{b}}$, where $\boldsymbol{F}_{i \to \mathrm{b}}$ is the contribution to the drag force acting on the cell body by the flagella-induced flow, i.e., the flagella-to-body (F-to-B) force. Using the strength of the image Stokeslet in $\mathsfbi{G}_{\mathrm{im}}$, $\boldsymbol{F}_{i \to \mathrm{b}}$ can be expressed as~\citep{Higdon79,Kim13}
\begin{equation}\label{f_to_b}
\boldsymbol{F}_{i \to \mathrm{b}} = -\left[\left(\frac{3 r_\mathrm{b}}{2R_i} - \frac{r_\mathrm{b}^3}{2 R_i^3} \right) \hat{\boldsymbol{R}}_i\hat{\boldsymbol{R}}_i + \left(\frac{3 r_\mathrm{b}}{4 R_i} + \frac{r_\mathrm{b}^3}{4 R_i^3}\right)(\mathsfbi{I} - \hat{\boldsymbol{R}}_i\hat{\boldsymbol{R}}_i)\right] \bcdot \boldsymbol{F}_i,
\end{equation}
where $\boldsymbol{R}_i = \boldsymbol{x}_i - \boldsymbol{x}_\mathrm{b}$ and $\hat{\boldsymbol{R}}_i = \boldsymbol{R}_i/|\boldsymbol{R}_i|$. The body velocity $\boldsymbol{U}_{\mathrm{b}}$ can be obtained from the force-free condition: $\boldsymbol{F}_\mathrm{b} + \sum_i \boldsymbol{F}_i = 0$. 

\subsection{Effective non-reciprocal interactions between cell body and flagella}
\begin{figure}
\centerline{\includegraphics{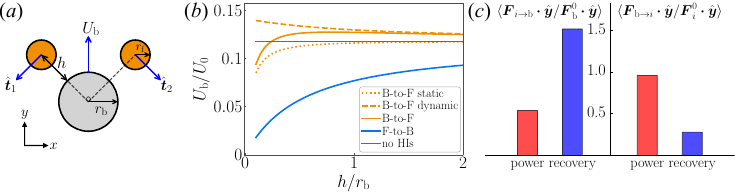}}
\caption{(\textit{a}) Schematic of the three-sphere model. (\textit{b}) Effect of different components of body-flagella HIs on the body speed $U_\mathrm{b}$ in the three-sphere model. The radius of the flagellar sphere is $r_\mathrm{f} = 0.1\ r_\mathrm{b}$. (\textit{c}) Average of the normalized drag forces $\langle \boldsymbol{F}_{i\to \mathrm{b}} \bcdot \hat{\boldsymbol{y}} / \boldsymbol{F}_\mathrm{b}^0 \bcdot \hat{\boldsymbol{y}}\rangle$ and $\langle \boldsymbol{F}_{\mathrm{b}\to i} \bcdot \hat{\boldsymbol{y}} / \boldsymbol{F}_i^0 \bcdot \hat{\boldsymbol{y}}\rangle$ (see text) over the power and recovery strokes.}
\label{fig4}
\end{figure}

The three-sphere model (\S~\ref{three_sphere}) enables us to separately evaluate the effects of the F-to-B force and the dynamic and static components of the B-to-F force. We consider the time instant when the three spheres form an isosceles right triangle. The left flagellar sphere moves in the direction $\hat{\boldsymbol{t}}_1 = (\cos\alpha, \sin\alpha, 0)$, where we set the angle relative to the $x$ axis as $\alpha = 5\pi/4$. Reversing the direction of motion by setting $\alpha = \pi/4$ corresponds to switching from the power stroke to the recovery stroke, and reverses the sign of $U_\mathrm{b}$. Figure~\ref{fig4}(\textit{b}) shows that the F-to-B force alone can significantly reduce $U_\mathrm{b}$ compared with that without any HIs. This implies that the reduction in swimming speed during the recovery stroke (figure~\ref{fig3}) results from the impedance of the cell body by the flagella-induced flow. The dynamic component of the B-to-F force increases $U_\mathrm{b}$, while the static component has the opposite effect but decays more rapidly with the distance from the body surface $h$. When $h \gtrsim r_\mathrm{b}$, which is approximately the average separation between the flagella and cell body (figure~\ref{fig1}\textit{a}), the static component has a negligible effect on $U_\mathrm{b}$. Therefore, the slight increase in $U_\mathrm{b}$ during the power stroke (figure~\ref{fig3}\textit{a}) may arise from the dominance of the dynamic component of the B-to-F force.

To verify the predictions from the three-sphere model, we evaluate how the cell body and flagella contribute to the drag forces on each other in the flagellated swimming model (\S~\ref{model}). We approximate the spheroidal cell body as a sphere of radius $r_{\mathrm{b}} = (a+b)/2$ and compute the F-to-B force $\boldsymbol{F}_{i\to \mathrm{b}}(t)$ by integrating (\ref{f_to_b}) along the arclength $s$, with $\boldsymbol{F}_i$ replaced by the flagellar force density $\boldsymbol{f}_i(s)$. The dynamic component of the B-to-F force is computed as
\begin{equation}
\boldsymbol{F}_{\mathrm{b} \to i}(t) = \int_{C_i} [\zeta_\parallel \boldsymbol{p}_i\boldsymbol{p}_i + \zeta_\perp(\mathsfbi{I} - \boldsymbol{p}_i\boldsymbol{p}_i)] \bcdot \boldsymbol{v}_\mathrm{b}[\boldsymbol{x}_i(s,t), \boldsymbol{x}_{\mathrm{b}}; \boldsymbol{U}_{\mathrm{b}}, r_\mathrm{b}] \, \mathrm{d}s,
\end{equation}
where $\zeta_\parallel$ and $\zeta_\perp$ are the parallel and perpendicular drag coefficients of the slender flagella~\citep{Cox1970}, and $\boldsymbol{p}_i$ is the unit tangent vector. To compare the relative strength of the effects of $\boldsymbol{F}_{i\to \mathrm{b}}$ on the cell body and $\boldsymbol{F}_{\mathrm{b} \to i}$ on the flagella, we normalize $\boldsymbol{F}_{i\to \mathrm{b}}$ by $\boldsymbol{F}_\mathrm{b}^0(t) = -6 \pi\mu r_\mathrm{b} \boldsymbol{U}_\mathrm{b} (t)$, and $\boldsymbol{F}_{\mathrm{b} \to i}$ by
\begin{equation}
\boldsymbol{F}_i^0 (t) = -\int_{C_i} [\zeta_\parallel \boldsymbol{p}_i\boldsymbol{p}_i + \zeta_\perp(\mathsfbi{I} - \boldsymbol{p}_i\boldsymbol{p}_i)] \bcdot [\boldsymbol{U}_\mathrm{b}(t) + \boldsymbol{U}_i (s, t)] \, \mathrm{d}s.
\end{equation}
Here, $\boldsymbol{F}_\mathrm{b}^0$ and $\boldsymbol{F}_i^0$ are the drag forces on the cell body and flagella, respectively, due to their own translational motions in a quiescent fluid. We normalize the components of $\boldsymbol{F}_{i\to \mathrm{b}}$ and $\boldsymbol{F}_{\mathrm{b} \to i}$ along the swimming direction $\hat{\boldsymbol{y}}$, and compute the average over the power and recovery strokes separately. 

Figure~\ref{fig4}(\textit{c}) shows that the body-flagella HIs are effectively nonreciprocal during both the power and recovery strokes. During the power stroke, $\langle \boldsymbol{F}_{\mathrm{b}\to i} \bcdot \hat{\boldsymbol{y}} / \boldsymbol{F}_i^0 \bcdot \hat{\boldsymbol{y}}\rangle$ is larger than $\langle \boldsymbol{F}_{i\to \mathrm{b}} \bcdot \hat{\boldsymbol{y}} / \boldsymbol{F}_\mathrm{b}^0 \bcdot \hat{\boldsymbol{y}}\rangle$, and the effect of the B-to-F force is stronger than that of the F-to-B force, leading to a slightly larger forward speed (figure~\ref{fig3}\textit{a}). During the recovery stroke, $\langle \boldsymbol{F}_{i\to \mathrm{b}} \bcdot \hat{\boldsymbol{y}} / \boldsymbol{F}_{\mathrm{b}}^0 \bcdot \hat{\boldsymbol{y}}\rangle$ is larger than $\langle \boldsymbol{F}_{\mathrm{b}\to i} \bcdot \hat{\boldsymbol{y}} / \boldsymbol{F}_i^0 \bcdot \hat{\boldsymbol{y}}\rangle$, and the effect of the F-to-B force dominates, leading to a smaller backward speed. The reversal of non-reciprocity when switching from the power stroke to the recovery stroke (figure~\ref{fig4}\textit{c}) arises from the time-nonreciprocal deformation of the flagella. During the power stroke, the flagella are extended straight and oriented perpendicular to the swimming direction, resulting in a stronger B-to-F effect. During the recovery stroke, the flagella are coiled, more aligned with the swimming direction, and positioned closer to the cell body, resulting in a stronger F-to-B effect. 

\section{Conclusion}
We modeled the swimming dynamics of \textit{C. reinhardtii} using numerical simulations based on a boundary element method and experimentally measured flagellar waveform. We find that the swimming performance is significantly enhanced by the body-flagella HIs. As the cell body size increases, an optimal body size for swimming appears due to the competition between enhanced body-flagella HIs and increased viscous drag on the cell body. The measured average body size of \textit{C. reinhardtii} is close to the computed optimal value that maximizes the swimming efficiency. Using the three-sphere model, we demonstrate that the body-flagella HIs are effectively non-reciprocal: the body affects the flagella more strongly than vice versa during the power stroke, while the reverse holds during the recovery stroke. As a result, the forward speed is larger than that without HIs during the power stroke, and the backward speed is smaller during the recovery stroke. 

Our results provide a hydrodynamic interpretation for the characteristic body size of microalgae. The existence of an optimal cell body size also suggests that the relation between the swimming performance and cargo size may be non-monotonic for certain swimmers composed of multiple interacting components, such as the three-sphere swimmer which moves through sequential changes in the linear separations between the spheres~\citep{Nasouri2019}. This optimality could inform laboratory designs of biohybrid microrobots~\citep{Zhang2024}. 

Non-reciprocity generally emerges in interactions mediated by velocity fields. Another example of non-reciprocal HIs is found in dense cilia arrays~\citep{Hickey2023}. Two other geometric features of eukaryotic microorganisms, namely the ratio of flagellar wave amplitude to wavelength and the body aspect ratio, also align with hydrodynamic predictions~\citep{Lisicki2024}. Finally, beyond hydrodynamic effects, various factors can shape cell morphology. In particular, biological constraints naturally impose a lower limit on body size, as essential organelles such as the nucleus and chloroplasts require a minimum volume to function. 

\backsection[Acknowledgements]{We thank Jun Zhang and Yi Man for helpful conversations.}

\backsection[Funding]{S.H. acknowledges support from the National Natural Science Foundation of China (NSFC grant no. 12504234) and the Fundamental Research Funds for the Central Universities. D.W. acknowledges support from NSFC grant no. 12574238. Z.L. acknowledges support from NSFC grant no. 12202438.}

\backsection[Declaration of interests]{The authors report no conflict of interest.}

\backsection[Data availability statement]{All the data supporting this work and the simulation codes that reproduce the dynamics of the rigid scallop model using both the hybrid BEM method and the nonlocal slender body theory are available at \url{https://github.com/shiyuanhu/microalgal_swimming}.}

\backsection[Author ORCIDs]{X. Hu, https://orcid.org/0009-0002-2474-1287; Z. Liu, https://orcid.org/0000-0002-3158-3902; D. Wei, https://orcid.org/0000-0002-6226-0639; S. Hu, https://orcid.org/0000-0002-8415-4263}

\appendix
\section{}\label{appA}
As a test problem for our implementation of the hybrid boundary element method (BEM) and regularized Stokeslet method, we consider a rigid scallop consisting of two slender rigid filaments hinged at a common end. This model does not swim on average, but oscillates. The orientation of the upper filament is prescribed as $\theta(t) = \theta_0[\sin(\omega t) + 1]$, where $\theta_0$ is the amplitude, and $\omega$ is the frequency. The lower filament is mirror-symmetric. 
\begin{figure}
\centerline{\includegraphics{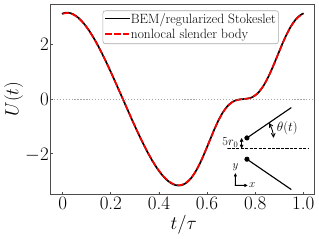}}
\caption{Comparison of the numerical results for a rigid scallop model, obtained from a nonlocal slender body theory and the hybrid boundary element and regularized Stokeslet method used in this work. The two filaments are separated by a small distance at the hinge point to avoid overlapping. The number of line elements $N=100$ and $m=6$.}
\label{figure_app}
\end{figure}

From a non-local slender body theory~\citep{Johnson1980}, the velocity of the upper filament $\partial \boldsymbol{x}_1/\partial t$ is related to the force densities $\boldsymbol{f}_i$ (for $i=1,2$) through 
\begin{equation}\label{nonlocal_slender}
8\pi\mu \left[\frac{\partial \boldsymbol{x}_1(s,t)}{\partial t} - \boldsymbol{v}_{2\to 1}(s,t)\right] = \left[c (\mathsfbi{I} + \boldsymbol{p}_1\boldsymbol{p}_1) + 2(\mathsfbi{I} - \boldsymbol{p}_1\boldsymbol{p}_1)\right]\bcdot \boldsymbol{f}_1(s) + \mathsfbi{K}[\boldsymbol{f}_1](s),
\end{equation}
where the slenderness parameter is $c = |\ln(\delta^2 e)|$, with $\delta$ the aspect ratio, the unit tangent vector is $\boldsymbol{p} = (\cos\theta, \sin\theta)$, and the integral operator $\mathsfbi{K}$ captures the non-local effects within the filament. The HIs between the two filaments are captured by
\begin{equation}\label{filament_HIs}
\boldsymbol{v}_{2\to 1}(s) = \frac{1}{8\pi\mu}\int_0^L \frac{\mathsfbi{I} + \hat{\boldsymbol{R}}(s,s')\hat{\boldsymbol{R}}(s,s')}{R(s,s')} \bcdot \boldsymbol{f}_2(s')\,\mathrm{d}s',
\end{equation}
where $\boldsymbol{R}(s,s') = \boldsymbol{x}_1(s) - \boldsymbol{x}_2(s')$. The filament velocity can be decoupled into a translation of the hinged point with velocity $U(t)\hat{\boldsymbol{x}}$ and a rotation with angular velocity $\dot{\theta}$.

We discretize the filament centerline using Chebyshev points. The operator $\mathsfbi{K}[\boldsymbol{f}](s)$ is first regularized~\citep{Tornberg04} and then evaluated via Clenshaw–Curtis quadrature with spectral accuracy~\citep{Trefethen2000}. Equations~(\ref{nonlocal_slender}) and (\ref{filament_HIs}) form a dense linear system for the unknowns $\boldsymbol{f}_i$ and $U(t)$. Figure~\ref{figure_app} shows that the result obtained from the hybrid BEM method using a line distribution of regularized Stokeslets agrees closely with the non-local slender body theory.

\section{}\label{appB}
\begin{table}
\begin{center}
\begin{tabular}{ccccc}
   Parameter  & Dataset 1 & Dataset 2 & Dataset 3 & Dataset 4\\
   \hline
   curvature $C$ (rad/$\mu$m) & $-0.11$ & $-0.12$ & $-0.13$ & $-0.17$ \\ 
   length $L$ ($\mu$m) &  13.97 & 10.82 & 13.03 & 9.42 \\
   wavelength $\lambda$ ($\mu$m) & 13.76 & 11.82 & 13.88 & 11.30 \\
   amplitude $A$ (rad) & 1.09 & 1.03 & 1.02 & 1.02 \\
   frequency $f$ (Hz) & 49 & 45 & 43 & 47 \\
   semi-minor $a$, semi-major axis $b$ ($\mu$m) & 3.89, 5.95 & 4.24, 4.61 & 3.91, 4.44 & 3, 3.53 \\
\end{tabular}
\caption{Variations in experimental waveforms explored in simulations. Dataset 1 is used in the main text.}
\label{table2}
\end{center}
\end{table}
\begin{figure}
\centerline{\includegraphics{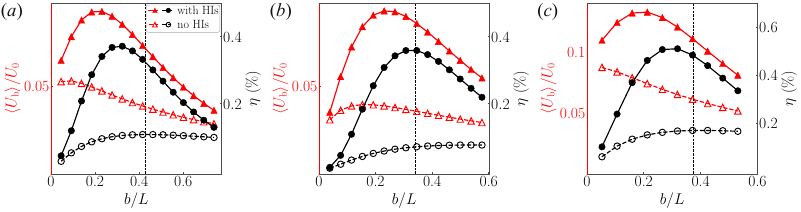}}
\caption{Effect of cell body size $b/L$ on the swimming speed $\langle U_\mathrm{b}\rangle/U_0$ and efficiency $\eta$, obtained from three additional waveforms measured in experiments (table~\ref{table2}): (\textit{a}) dataset 2, (\textit{b}) dataset 3, (\textit{c}) dataset 4. The vertical dashed lines mark the values of $b/L$ of each sample cell.}
\label{fig6}
\end{figure}

We quantify the geometric properties of waveforms used in our simulations by separating the tangent angle $\theta(s,t)$ into static and dynamic components~\citep{Geyer2016}. The static component is defined as the time average of the tangent angle over a beating period, $\bar{\theta}(s) = (1/T)\int_0^T \theta(s,t)\,dt$, and the dynamic component is defined as $\tilde{\theta}(s,t) = \theta(s,t)-\bar{\theta}(s)$. The static component $\bar{\theta}(s)$ is approximately linear in $s$, which corresponds to a circular shape. This static shape is characterized by its curvature $C = |\bar{\theta}(L) - \bar{\theta}(0)|/L$. The dynamic component $\tilde{\theta}(s,t)$ is close to a traveling wave with amplitude $A$ and wavelength $\lambda$. These geometric parameters are summarized in table~\ref{table2}. Although the four waveforms differ significantly in curvature $C$, length $L$, and wavelength $\lambda$, the computed swimming speed $\langle U_\mathrm{b} \rangle$ and efficiency $\eta$ consistently display pronounced optima as a function of the body size $b$ when HIs are incorporated (figure~\ref{fig6}).

\bibliographystyle{jfm}
\bibliography{jfm}

\end{document}